\documentclass[superscriptaddress,amsmath,amssymb,
aps,prl,twocolumn,noeprint,floatfix]{revtex4-1}
\usepackage{graphicx}
\usepackage{dcolumn}
\usepackage{bm}
\usepackage{hyperref}
\usepackage{notes2bib}
\usepackage{siunitx}
\usepackage{wasysym}
\usepackage[usenames,dvipsnames]{xcolor}
\usepackage{nicefrac}
\usepackage[caption=false]{subfig}
\definecolor{PR}{rgb}{0.57, 0.36, 0.51}
\definecolor{todo}{rgb}{1, 0.75, 0.0}
\definecolor{Sergey}{rgb}{0.2, 0.4, 0.8}
\definecolor{check}{rgb}{0.8, 0.4, 0.1}

\DeclareMathOperator{\sech}{sech}

\renewcommand{\l}{\ell}
\newcommand{\sigmaxx}{\sigma_\mathrm{xx}}
\newcommand{\sigmaxxavg}{\left<\sigmaxx\right>}
\newcommand{\sigmaxy}{\sigma_\mathrm{xy}}

\newcommand{\ABZ}{A_\text{osc}}
\newcommand{\A}{A}
\newcommand{\cA}{\mathcal{A}}

\renewcommand{\r}{{\mathfrak r}}

\usepackage[T1]{fontenc}

\newcommand{\R}{{\overset{\curvearrowright}{S}}}
\newcommand{\T}{{\overset{\curvearrowleft}{S}}}
\renewcommand{\L}{{\mathcal L}}

\newcommand{\beq}{\begin{equation}}
\newcommand{\eeq}{\end{equation}}
\newcommand\beqa{\begin{eqnarray}}
\newcommand\eeqa{\end{eqnarray}}
\newcommand\bea{\begin{array}}
\newcommand\eea{\end{array}}
\newcommand\ba{\begin{array}}
\newcommand\ea{\end{array}}
\newcommand{\nn}{\nonumber}
\newcommand{\eq}[1]{Eq.(\ref{#1})}

\newcommand{\Eq}[1]{Eq.(\ref{#1})}

\renewcommand{\d}{\partial}
\newcommand{\eV}{{\,\mathrm {eV}}}
\newcommand{\nm}{{\,\mathrm {nm}}}

\newcommand{\meV}{{\mathrm{\,meV}}}

\newcommand{\eps}{\epsilon}
\newcommand{\bmul}{\begin{multline}}
\newcommand{\emul}{\end{multline}}
\newcommand\comment[1]{}

\begin{document}

\title{
Kagom\'e quantum  oscillations in graphene superlattices
}

\author{Folkert K. de Vries}
\affiliation{Laboratory for Solid State Physics, ETH Z\"{u}rich, CH-8093 Z\"{u}rich, Switzerland}
\author{Sergey Slizovskiy}
\affiliation{National Graphene Institute, University of Manchester, Manchester M13 9PL, United Kingdom}
\affiliation{Department of Physics \& Astronomy, University of Manchester, Manchester M13 9PL, United Kingdom}
\email{Second author contributed equally} 
\author{Petar Tomić}
\affiliation{Laboratory for Solid State Physics, ETH Z\"{u}rich, CH-8093 Z\"{u}rich, Switzerland}
\author{Roshan Krishna Kumar}
\affiliation{National Graphene Institute, University of Manchester, Manchester M13 9PL, United Kingdom}
\affiliation{Department of Physics \& Astronomy, University of Manchester, Manchester M13 9PL, United Kingdom}
\affiliation{ICFO-Institut de Ciencies Fotoniques, The Barcelona Institute of Science and Technology, Barcelona, Spain}
\author{Aitor Garcia-Ruiz}
\affiliation{National Graphene Institute, University of Manchester, Manchester M13 9PL, United Kingdom}
\affiliation{Department of Physics \& Astronomy, University of Manchester, Manchester M13 9PL, United Kingdom}
\author{Giulia Zheng}
\author{El\'{i}as Portol\'{e}s}
\affiliation{Laboratory for Solid State Physics, ETH Z\"{u}rich, CH-8093 Z\"{u}rich, Switzerland}
\author{Leonid A. Ponomarenko}
\affiliation{Department of Physics, University of Lancaster, Lancaster LA1 4YW, United Kingdom}
\author{Andre K. Geim}
\affiliation{National Graphene Institute, University of Manchester, Manchester M13 9PL, United Kingdom}
\affiliation{Department of Physics \& Astronomy, University of Manchester, Manchester M13 9PL, United Kingdom}
\author{Kenji Watanabe}
\affiliation{National Institute for Materials Science, 1-1 Namiki, Tsukuba 305-0044, Japan}
\author{Takashi Taniguchi}
\affiliation{National Institute for Materials Science,  1-1 Namiki, Tsukuba 305-0044, Japan}
\author{Vladimir Fal'ko}
\email{vladimir.falko@manchester.ac.uk}
\affiliation{National Graphene Institute, University of Manchester, Manchester M13 9PL, United Kingdom}
\affiliation{Department of Physics \& Astronomy, University of Manchester, Manchester M13 9PL, United Kingdom}
\affiliation{Henry Royce Institute for Advanced Materials, M13 9PL, Manchester, United Kingdom}
\author{Klaus Ensslin}
\email{ensslin@phys.ethz.ch}
\affiliation{Laboratory for Solid State Physics, ETH Z\"{u}rich, CH-8093 Z\"{u}rich, Switzerland}
\author{Thomas Ihn}
\affiliation{Laboratory for Solid State Physics, ETH Z\"{u}rich, CH-8093 Z\"{u}rich, Switzerland}
\author{Peter Rickhaus}
\affiliation{Laboratory for Solid State Physics, ETH Z\"{u}rich, CH-8093 Z\"{u}rich, Switzerland}

\date{\today}
\maketitle

{\bf  Periodic systems feature the Hofstadter butterfly spectrum produced by Brown--Zak minibands of electrons formed when magnetic field flux through the lattice unit cell is commensurate with flux quantum and manifested by magneto-transport oscillations. Quantum oscillations, such as Shubnikov -- de Haas effect and Aharonov--Bohm effect, are also characteristic for electronic systems with closed orbits in real space and reciprocal space. Here we show the intricate relation between these two phenomena by tracing quantum magneto-oscillations to Lifshitz transitions in graphene superlattices, where they persist even at relatively low fields and very much above liquid-helium temperatures. The oscillations originate from Aharonov--Bohm interference on cyclotron trajectories that form a kagom\'e-shaped network characteristic for Lifshitz transitions. In contrast to Shubnikov - de Haas oscillations, the kagom\'e oscillations are robust against thermal smearing and they can be detected even when the Hofstadter butterfly spectrum is undermined by electron's scattering. We expect that kagom\'e quantum oscillations are generic to rotationally-symmetric two-dimensional crystals close to Lifshitz transitions.}

\begin{figure}[t]
\includegraphics[width=0.9\columnwidth]{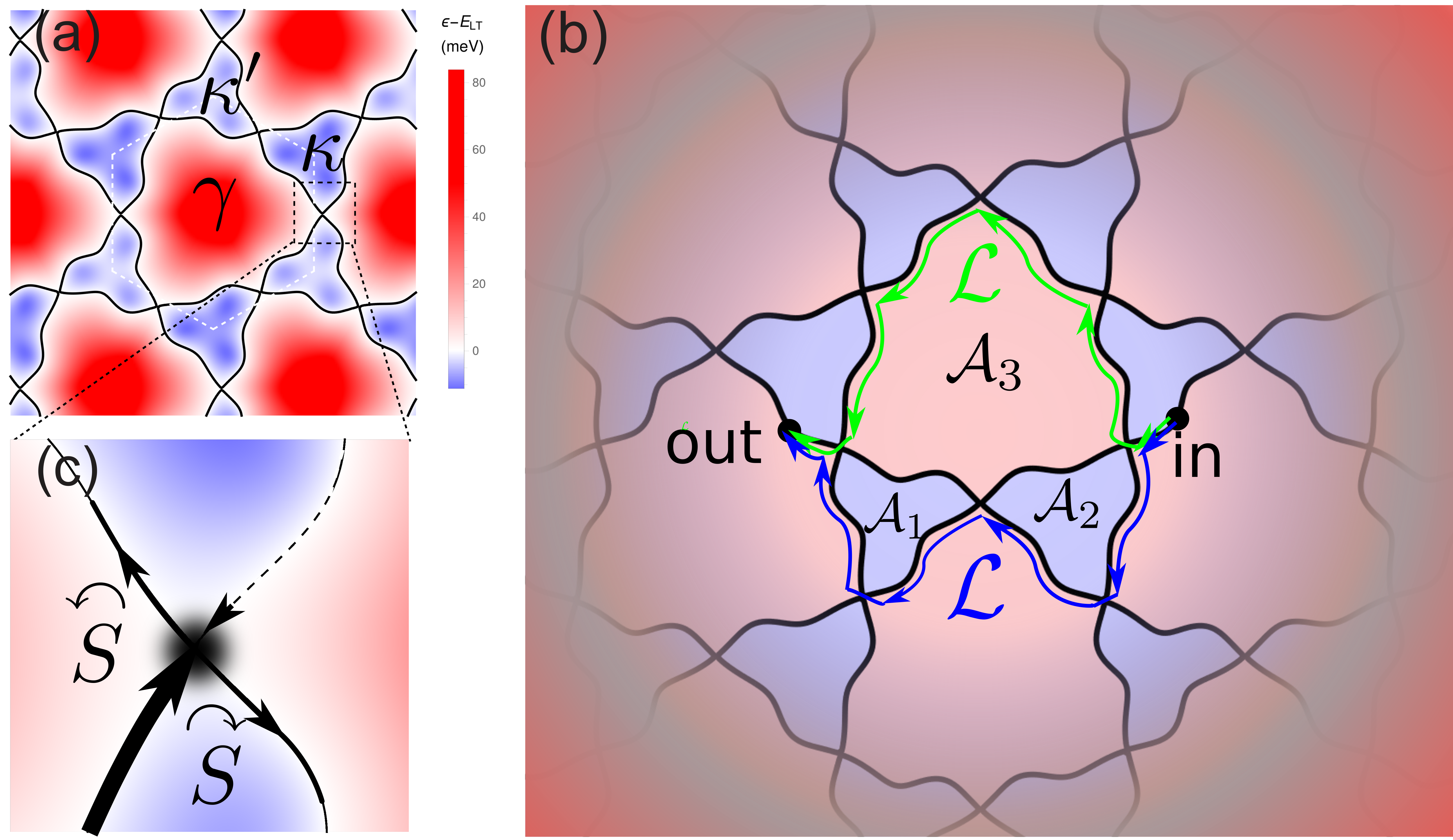}
\caption{(a) Fermi contours for LT and dispersion map for the 1$^{st}$ miniband in the K-valley conduction band of a AB/BA tDBLG (modelled using Hamiltonian described in Ref. \cite{AitorTrigonal} for $\theta =1.9^\circ$ twist angle); $\kappa$, $\kappa'$ and $\gamma$ mark the high-symmetry points in the mini Brillouin zone of mSL. Red/blue indicate miniband energies above/below $E_{\text{LT}}$. (b) At the LT, ballistic trajectories of electrons in a magnetic field form a trihexagonal kagom\'e network. Green and blue lines in (b) exemplify the shortest paths responsible for quantum magneto-oscillations at the LT, zoomed-in for $E=E_{\text{LT}}+\epsilon$. Note that both having the same length, $\L$. (c) Electrons scatter at the intersections of paths linked to the saddle points in the band dispersion. }
\label{fig:fig1} 
\end{figure}

Lifshitz transitions~\cite{Lifshitz1960} (LTs) are generic for the electronic bands in solids. They mark the sign change for the effective mass of electrons, accompanied by saddle-point features in electrons' dispersion and van Hove singularities in their density of states. In two-dimensional (2D) crystals, a LT also singles out a band energy for which disconnected closed-loop contours, $\varepsilon ({\bf p})=\rm const$, merge into a multiply-connected network. For most generic LTs where the 'electron-like' dispersion transforms into 'hole-like' near its top, these constant-energy contours, $\varepsilon ({\bf p})= E_{\text{LT}}$, form networks, spanning across the entire reciprocal space of a crystal (see an example in Fig.~\ref{fig:fig1}a).

Constant energy maps are important for  magnetotransport phenomena, as they fully determine the shape of ballistic electron trajectories in a 2D metal subjected to magnetic field, ${\bf B} = B \hat{\bf z}$. Because the  electron's dynamics is set by $\dot {\bf p} = eB \hat{\bf z} \times \dot {\bf r}$ and $\dot {\bf r} \equiv {\bf v} = \nabla_{{\bf p}} \eps({\bf p})$, the real-space trajectories can be obtained from constant-energy contours by a $90^\circ$ rotation and re-scaling using a $(eB)^{-1}$ factor, Fig.\,\ref{fig:fig1}b. For closed-loop energy contours, this transformation results in closed cyclotron trajectories. Aharonov--Bohm interference along such trajectories is known to lead to semiclassical Shubnikov--de Haas oscillations (SdHO) that are forerunners of the fully quantized electronic spectrum and the appearance of, for example, the quantum Hall effect. In this report, we show that multiply connected trajectories near Lifshitz transitions  ($\epsilon \approx E_{\text{LT}}$), lead to a peculiar interference contribution to conductivity which turns out to be insensitive to thermal broadening of the Fermi step (unlike SdHO) due to the trihexagonal geometry of these trajectories (Fig.~\ref{fig:fig1}b) ), whereas the magnetic breakdown spans the oscillations over a substantial interval of carrier densities around LTs. We refer this behaviour as kagom\'e oscillations, as the low-field forerunners of Brown--Zak minibands \cite{Brown1964, Zak1964, Hofstadter1976} which determine the transport properties of the latter. Manifestations of the latter in quantum oscillations of capacitance \cite{Ponomarenko2013} and transport \cite{Dean2013, Kumar2017, Saito2021} have been observed in several graphene superlattices at high magnetic fields, and they were even seen to persist at high-temperatures \cite{Kumar2017, Kumar2018}. Here, we show that the high-temperature magneto-transport oscillations emerge from kagom\'e oscillations near Lifshitz transitions in graphene superlattices.  

\begin{figure*}[h!]
\includegraphics[width=0.9\textwidth]{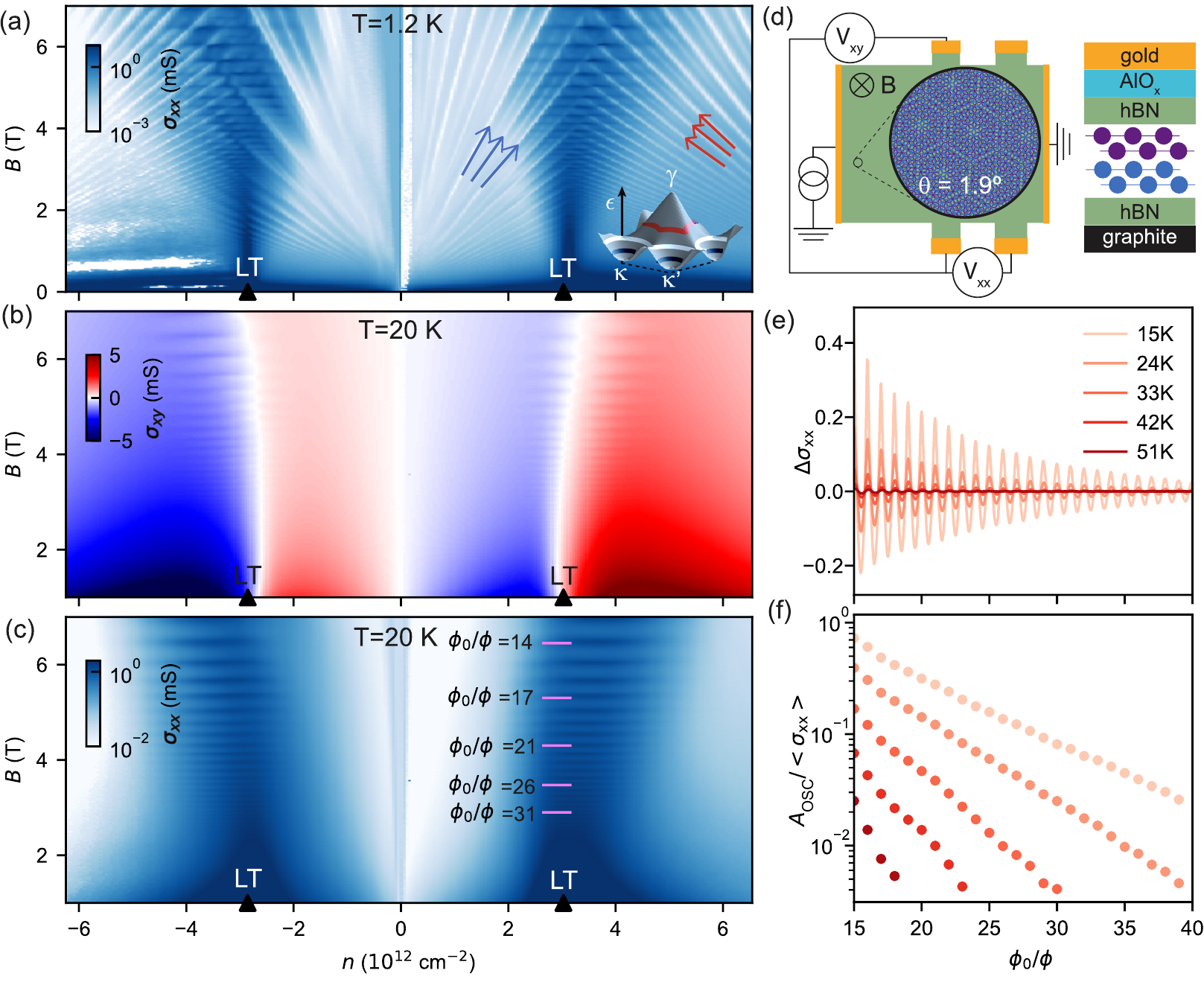}
\caption{Magneto-transport oscillations in a tDBLG device with $\theta=1.9^\circ$, sketched in (d).
(a) Landau fan diagram $\sigmaxx(n,B)$ at $T=\SI{1.2}{K}$. Arrows mark Landau levels formed around $\kappa$'s (blue) and $\gamma$ (red) edges of the 1$^{st}$ miniband on the conduction band side of graphene, shown in the inset. (b) The LTs are identified from the changing sign of the low-field $\sigmaxy$. (c) $\sigmaxx(n,B)$ at $\SI{20}{K}$. Oscillations, periodic in  $\phi_0/\phi$, are pronounced in the vicinity of the LTs. (d) Schematic representation of the tDBLG device; its double-gating enables tuning the displacement field to $D=0$ for each electron density $n$. (e) Oscillating part of conductivity, $\Delta\sigmaxx$ at $n=\SI{3.2e12}{cm^{-2}}$, and (f) the amplitude of these oscillations, plotted as a function of $\phi_0/\phi$ for various temperatures using colour coding from (e) (see SM \cite{SM} for details). }
\label{fig:fig2}
\end{figure*}

In Fig.~\ref{fig:fig1}a and Fig.~S5 we illustrate a characteristic LT contour for lattices which possess a C$_3$ rotational symmetry. The illustrated geometry of the LT contour is generic for graphene minibands formed by moir\'e superlattices (mSL) in graphene-hBN heterostructures (G/hBN)  ~\cite{Yankowitz2012,Ponomarenko2013,Kumar2017,Kumar2018}, twisted graphene bilayers ~\cite{Lin2020,shen2021}, trilayers  ~\cite{Trilayer12,Chen16trilayer,TwistReview}, and double-bilayers   ~\cite{Burg2019,Liu2020,Shen2020,He2021} (tDBLG). 
It has the distinct form of a trihexagonal kagom\'e network (which would be inverted for graphene's K and K' valleys), with a hexagonal part centered at $\gamma$ and two triangular shapes around $\kappa$ and $\kappa'$ points of the mini Brillouin zone of mSL. The corresponding hexagonal and triangular areas in Fig.\,\ref{fig:fig1}a are painted in red and blue, indicating which parts of a particular miniband dispersion are at the energies above and below $E_{\text{LT}}$, respectively. The example shown in Fig.\,\ref{fig:fig1} corresponds to the 1$^{\text{st}}$ mSL miniband on the conduction band side of tDBLG spectrum (to represent the LT map for the  1$^{\text{st}}$ mSL miniband on the valence band side, we would have to swap red and blue colours). 

Below, we relate the kagom\'e-shaped LT contours to magneto-transport oscillations observed in different graphene-based superlattices. We start by discussing the oscillations observed in a tDBLG, Fig.~\ref{fig:fig2}. These oscillations were measured in a Hall bar device \cite{SM,Tomic2021}, Fig.~\ref{fig:fig2}d, at low magnetic fields and in the \SI{10}{K}--\SI{50}{K} temperature range, where they replace SdHO dominant at low temperatures, Figs.~\ref{fig:fig2}a. The latter, reflecting Landau levels (LL) formed by electrons revolving along closed cyclotron loops (Fig.S1%\ref{suppfig:LLfits} 
), have an amplitude, $\Delta\sigma_\text{SdHO} \propto \sech \frac{2\pi^2 k_B T\rho}{B / \phi_0}$, suppressed at temperatures larger than the LL splitting \cite{LifshitsKosevich}, in particular, around LTs, where the density of states of electrons, $\rho$, is high. The limitations on the appearance of SdHO and the data on the measured transport mean free path, $\ell_\text{mfp}$ in the studied tDBLG are discussed in Figs. S7 and S8. The remaining high-temperatures oscillations have $\sigma_{xx}$ maxima when the magnetic flux through the mSL unit cell, $\phi = B\cA_{\varhexagon}$, is commensurate, $\phi = \phi_0/q$ with an integer $q$, with the flux quantum, $\phi_0 =h/e$.  While this formally coincides with the conditions for Brown--Zak minibands formation ~\cite{Brown1964,Zak1964,Hofstadter1976}, the oscillations in Fig.~\ref{fig:fig2}c are clearly visible at temperatures largely exceeding the miniband widths and gaps in-between them. Below, we argue that these oscillations are produced by the interference contribution to the electronic transport which would be present (though, in a weaker form) even if the length of the electron mean free path, $\ell_\text{mfp}$ would be comparable to the perimeter, $\L_q \sim 4\sqrt{\frac{2}{\sqrt{3}}} q \sqrt{\cA_{\varhexagon}}$, of a magnetic supercell \cite{Brown1964,Zak1964,Chen2014}. 

Let us consider an unbound propagation of an electron along the kagom\'e network at the LT in Fig.~\ref{fig:fig1}b. Due to the quantum nature of electrons, their propagation along such a network has a stochastic element. The meander-like paths bifurcate at their intersections, Fig.~\ref{fig:fig1}c, associated with the saddle-point features in the electron dispersion, captured by transmission amplitudes, $\R$ and $\T$ \cite{Glazman18, Berry72, Wilkinson, Davis}, 
\beq
\label{RT}
\begin{split}
&\frac{|\R|}{|\T|} = e^{\pi \mu};  \,   
\mu = \frac{\hbar\, (\eps-E_{\text{LT}})}{e B\r}; \, |\R|^2+|\T|^2=1;  
\\
 &\arg \T = \arg\R = \arg\left[\Gamma\left(\frac12-i \mu\right)\right] + \mu (\ln|\mu| - 1). 
\end{split}
\eeq
These amplitudes are comparable within the 'magnetic breakdown' \cite{Cohen1961} energy interval $\sim eB\r/h$, dependent on the Gaussian curvature of the dispersion saddle-point, $\r = \hbar \sqrt{|\det \frac{\d^2 \eps ({\bf p})}{\d {p_i} \d {p_j}}|}$, which determines the energy window around $E_{\text{LT}}$ where the LT kagom\'e network remains relevant for  electron transport. This makes the LT network similar to the networks of topologically protected channels at the AB/BA domain boundaries in marginally twisted bilayer graphene \cite{HelicalStatesExpt1,berdyugin2019}. However, the LT networks of trajectories --- such as in Fig. \ref{fig:fig1}b --- are not set in stone, as each of those is linked to the initial position and velocity, $|in \rangle$, of an electron propagating ballistically. 
%being on a ballistic propagation.

Having in mind that the lengths of electronic trajectories, $\L \propto B^{-1}$, scale up and can exceed the electron mean free path at $B\rightarrow 0$, in Fig. \ref{fig:fig1}b we exemplify the shortest paths visiting several LT network nodes on the way from point 'in' to 'out'. Interference of electron waves propagating along those paths contributes towards the probability for an electron to reach 'out' from 'in'. The phases, acquired by an electron along this pair of paths, have the same property, as noted in an AB/BA domain wall network \cite{tBLGAB}: as these two paths are composed of pairs of equivalent ballistic segments, the electron-energy-dependent dynamical part of the phase related to the length of the path is the same for both of them, making their interference condition (constructive or destructive) set only by the encircled magnetic field flux, independent of the electron's energy deviation from $E_{\text{LT}}$. Note that the above-mentioned cancellation of dynamical phases is also similar to the phase cancellations for selected pairs of paths in the weak localisation effect in electronic systems. It leads to the resilience of the resulting interference effects against thermal broadening of the Fermi step. 

\begin{figure*}[h!]
\centering
\includegraphics[width=0.9\textwidth]{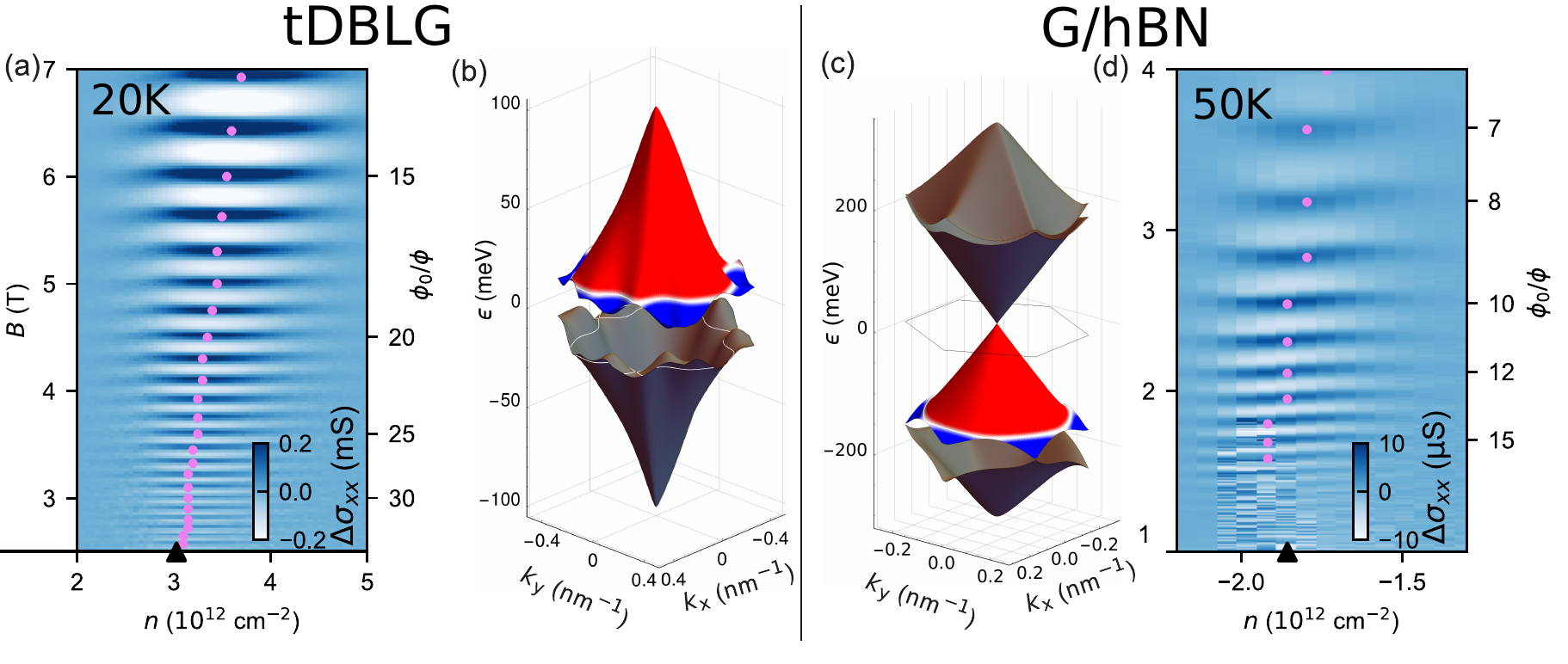}
\caption{
Comparison of $\sigma_{xx}$ oscillations around LTs observed in moire superlattice minibands in tDBLG and G/hBN. Dots mark the shift of the maxima of oscillations' amplitudes with increasing magnetic field. 
(a) Oscillations of $\sigma_{xx}(q,n)$ around LT in the first miniband on the conduction band side of tDBLG spectrum, shown in (b). The experimental data are plotted after subtracting a smooth background from data in Fig. 2 (see \cite{SM} and Figs. S2,S3 for details); T = 20K. (c)  Electronic spectrum of G/hBN superlattices \cite{Yu2014,LeeWallbank16}) which features a clear LT in the 1st miniband on the valence band side of graphene's spectrum (white contour). In agreement with Eq.\,(\ref{oscillation_main}), similarity between the highlighted minibands in (b) and (c) prescribes the same direction of a shift (upon adding electrons to the system) of the oscillations maxima from the LT in tDBLG and (d) in a highly aligned G/hBN device, where mSL has a $15$\,nm period (here, $\sigma_{xx}$ was measured at $T=\SI{50}{K}$ and analysed as in tDBLG device.}
\label{fig:fig3}
\end{figure*}

To take into account interference of these two highlighted partial waves, we write an amplitude for an electron to reach 'out' from 'in', as 
\beq
\label{amplitude}
\langle {\rm out |in }\rangle = \alpha_\text{diff} + e^{i \varphi} \left[ \T^3 \R^2 + \T^3 \R^2 e^{i \frac{ e B \cA}{\hbar}} \right] e^{-\frac{\L}{\l}}.
\eeq
Here, $\varphi =  \frac{1}{\hbar}\int_\text{in}^\text{out} \bf p \cdot dr$ is the phase acquired by an electron propagating along the green path in Fig.\,\ref{fig:fig1}b, and   
$$\cA \equiv \cA_1+ \cA_2 + \cA_3 =\frac{\A_\text{BZ}}{(e B)^2} \equiv \frac{\phi_0^2}{\cA_{\varhexagon} B^2},$$
is the area between the green and the blue paths that is equal to the area of the kagom\'e network unit cell ($\A_\text{BZ}$ and $\cA_{\varhexagon}$ are the mSL Brillouin zone area and the unit cell area, respectively). We also account for Maslov's phase multipliers $\mp i$ for each clockwise/anticlockwise turning point \cite{SM,MaslovTopological,Glazman18} and list all the shortest interfering paths in Fig. S4. 
%(see SI for details). 
An amplitude $\alpha_\text{diff}$ accounts for diffusive paths with variable lengths linking "in" to "out" upon scattering from graphene's disorder, or phonons, which may even be the dominant transport mechanism. Such scattering processes are also accounted for by a factor $e^{-\L/\l}<1$, and we neglect contributions of longer inter-crossing ballistic paths, as their contributions would be additionally suppressed by disorder scattering. 

For the probability, $|\langle {\rm out |in }\rangle |^2 $, of an electron to reach 'out' from 'in', the interference of partial waves arriving along such pairs of paths is set by the Aharonov--Bohm phase, $eB \cA/\hbar$. This produces a contribution,
\beq 
\label{Oscillations0}
\Delta \sigma_{\text{LT}}  \sim  \frac{e^2}{h}  \cos \frac{2 \pi \phi_0}{\cA_{\varhexagon} B} \times e^{-\frac{2\L}{\l}} \int d\eps \, \frac{\d n_F}{\d \eps}  |\T^6 \R^4|,
\eeq
to the conductivity $\sigma_{xx}=\sigma_0+\Delta \sigma_{\text{LT}}$, which oscillates periodically as a function of $B^{-1}$. This contribution comes on top of a diffusive conductivity background, $\sigma_0$, in which the interference between waves following paths of variable long lengths, with large energy-dependent phases, is washed out by averaging over disorder and over energies, due to the thermal broadening of the Fermi step (here, accounted for by the factor $\frac{\d n_F}{\d \eps}$). 

Equation (\ref{Oscillations0}) describes the high-temperature low-field $1/B$-periodic oscillations, Fig. \ref{fig:fig2}c, observed in tDBLG around $n_{\text{LT}}$ (here, LT was identified by the sign change of the lowest-field Hall conductivity, Fig. \ref{fig:fig2}b). In contrast to SdHO, shown in Fig. 2a for 1.2\,K but disappearing above liquid-helium temperature, kagom\'e oscillations sustain elevated temperatures (up to 50\,K, Fig.\,\ref{fig:fig2}e) and have a period set by the unit cell area of mSL, independent of the gate-induced carrier density, $n$. 

We note that the observed kagom\'e oscillations not only appear at temperatures largely exceeding Brown--Zak miniband widths and gaps \cite{Brown1964,Zak1964,Hofstadter1976,Kumar2017,Kumar2018}, but, also, are noticeable even in the regimes of a substantial disorder and phonon scattering, $\ell_\text{mfp} \le 2\L_q$, where Brown--Zak minibands cannot form. To discuss the influence of scattering on kagom\'e oscillations, accounted for by the factor $e^{-2\L/\l}$ in Eq.\,(\ref{Oscillations0}), we analysed the envelopes of the oscillations in Fig.\,\ref{fig:fig2}e, which decay as $e^{-\beta/B}$ over two orders of magnitude, Fig.\,\ref{fig:fig2}f and Fig.~S6. Such exponential decay is a natural consequence of scaling, $\L(B) \propto B^{-1}$, of the kagom\'e network segments, and the exponential decay of the measure amplitudes of kagom\'e oscillations shown in Fig. 2 suggest that their observation is made in the scattering-dominated transport regime. 

The high-temperature oscillations in Fig.\,\ref{fig:fig2}c are the largest near the LTs (here, in the 1$^\text{st}$ miniband on both conduction and valence side of tDBLG spectrum, but their maxima shift away from $n_{LT}$ upon increasing magnetic field (this shift largely exceeds small deviations of the zero of $\sigma_{xy}$ from the vertical direction in Fig.\,\ref{fig:fig2}b, which can be attributed to anomalous contributions to the Hall conductivity induced by the Berry curvature). We relate this shift, highlighted in Fig.\,\ref{fig:fig3}a to the kagom\'e network topology and a typical --- for LTs in  all C$_3$-symmetric crystals --- energy dependence of transmission amplitudes in Eq.\,(\ref{Oscillations0}). Indeed, the interference term in Eq.\,(\ref{Oscillations0}) contains an unbalanced product of $|\R|^2$ and $|\T|^2$ factors, $\left |\T^6 \R^4\right |= \frac{e^{-\pi \mu}}{32 \cosh^5 \pi \mu}$, which has the energy dependence skewed towards lower energies. 
A saddle-point evaluation of the integral in Eq. (\ref{Oscillations0}), recalculated from the Fermi energy into carrier density dependence (using density of states in the band, $\rho$, averaged over $k_B T$ interval around LT), results in the following expression for the low-field tail of oscillations,  
\beq \label{oscillation_main}
 \begin{split}
 \Delta \sigma_\text{LT} \propto \frac{B}{\delta n}e^{-\frac{(n-n_{\max})^2}{\delta n^2}} 
e^{-\frac{2\L(B)}{\l(T)}}  \,
 \cos \frac{2 \pi \phi_0}{\cA_{\varhexagon} B}; \\
 n_{\max}= n_{\text{LT}}\pm \frac{2 \rho B\r}{5 \phi_0}; \, \,
 \delta n =2 \rho  \sqrt{(k_B T)^2+ \frac25 \left(\frac{B\r}{\phi_0}\right)^2 }.
\end{split}
\eeq
Here, $\pm$ accounts for the type of dispersion inside the hexagonal part of mSL Brilloun zone around $\gamma$: it is $+$ for hole-like ($\epsilon(\gamma) > E_{\text{LT}}$, as shown in Fig.\,\ref{fig:fig3}b) and $-$ for electron-like ($\epsilon(\gamma) < E_{\text{LT}}$) part of the miniband. As a result, the inverted form of minibands on the valence and conduction band side of tDBLG spectrum (see Fig.\,\ref{fig:fig3}b) sets the opposite shift directions, $n_{\text{LT}}\pm \frac{2 \rho \r B}{5 \phi_0}$, for the oscillations amplitude maxima near the LT in n- and p-doped tDBLG.

To corroborate the generality of the trend described by Eq.\,(\ref{oscillation_main}) for graphene and other C$_3$-symmetric superlattices, we compare tDBLG to monolayer graphene placed and aligned on hBN (G/hBN), where mSL is formed with a 15\,nm period \cite{Kumar2017}. The mSL minibands for such a system are shown in Fig.\,\ref{fig:fig3}c. Here, we stress that, for G/hBN, the hole-like dispersions around $\gamma$ are separated by the LT from electron-like pockets around $\kappa$'s in the 1$^{st}$ miniband on the valence band side in the same way as in the 1$^{st}$ miniband on the conduction band side in tDBLG. This similarity determines the same direction of the shift of the oscillations maxima from the respective LTs upon adding electrons in both systems, Fig.\,\ref{fig:fig3}a,d. The $\Delta\sigma _{xx}$ data for G/hBN were taken at $T=\SI{50}{K}$ and analysed for the low magnetic field end in the same way as for tDBLG (Fig.\,\ref{fig:fig3}d), whereas at high magnetic fields they agree with the earlier observed magneto-oscillations \cite{Kumar2017} which reflect recurrent appearance of faster-propagating \cite{Kumar2018} Bloch states at superlattice-commensurate flux values, $\phi_0/q$.   

Also, from the computed mSL minibands, we estimate $\r \approx 1.2\, \eV \nm^2$ for tDBLG and $\r \approx 6 \, \eV \nm^2$ for G/hBN ~\cite{SM}; as a result, we find that $\delta n$ is temperature-limited up to $B \sim 10$ T for tDBLG at $T = 20$ K and up to $B\sim 5$ T for G/hBN at $T=50$ K, explaining the weak $B$-dependence of the width of the regions where the oscillations occur in Fig. \ref{fig:fig3}a,d. Fitting the measured magnetic field dependence of oscillations amplitudes with \eq{oscillation_main} and length $\L$, Fig.S6
%\ref{suppfig:dephasing} 
%S6
, we find temperature dependence of  scattering parameter, $\ell^{-1}(T)$, and compare it, in   Fig.S7
%\ref{suppfig:mfpPlot}
, to the inverse of the measured mean free path, $\ell^{-1}_\text{mfp}(B=0,\varepsilon=E_{LT})$. This comparison shows that (a) momentum relaxation (which requires larger momentum transfer) is dominated by electron-phonon scattering in the Bloch-Grüneisen regime, and (b) that it is slower than decay of a ballistic wave packet, which could be expected as the latter is sensitive to the large momentum transfer upon scattering, whereas $\ell^{-1}$ is caused by both large and small angle scattering. After this, we note that formation of Brown--Zak minibands \cite{Brown1964,Zak1964} and Hofstadter butterfly are determined by Bragg scattering from the superlattice with a period inflated $q$ times as compared to the underlying moir\'e pattern \cite{Chen2014}. For forming minibands the interfering electron waves should propagate ballistically over - at least - two magnetic supercells (to explore periodicity of magnetic superlattice), which requires that $2\L_q(B) \le \ell_\text{mfp}(T)$, hence, higher magnetic fields. Figure S8
%\ref{suppfig:Bstar}
%S8
shows that the kagom\'e oscillations were observed in the opposite regime of low magnetic fields, heralding the formation of minibands and anticipating the oscillations of kinetic parameters (such as characteristic group velocities) simultaneously occuring in several consecutive Brown--Zak minibands) in the mSL spectrum. 

While the studies of kagom\'e oscillations in this paper were focused on moir\'e superlattices with trigonal (C$_3$) symmetry, we expect similar forerunner of Brown--Zak oscillations to exist in crystals with a C$_4$ symmetry \cite{Azbel1964,Hofstadter1976}, too. However, we note that they would be suppressed in systems with a lower symmetry. This is because LT contours in low-symmetry crystals have the form of quasi-1D block-chains of intertwining meanders \cite{Kaganov1979,Slutskin83} which do not provide pairs of paths needed for the energy-independent Aharonov--Bohm interference. Therefore, breaking the C$_3$ rotational symmetry of the mSL by straining one of the 2D crystals in a stack \cite{FaradayDiscussions} and violating the kagom\'e topology of the LT network would suppress these novel low-magnetic-field high-temperature quantum oscillations.

\section{Acknowledgements}
We acknowledge support from the European Graphene Flagship Core3 Project, EPSRC grants EP/W006502/1, EP/V007033/1, EP/S030719/1, Swiss National Science Foundation via NCCR Quantum Science, EU H2020 grant No 862660/QUANTUM E LEAPS, and H2020 European Research Council (ERC) Synergy Grant under Grant Agreement 951541.

\section{Author contributions}
K.E.,T.I. P.R. and V.F. conceived the work and designed the research strategy. F.K.deV. and P.T. performed conductance measurements and data analysis under K.E.’s supervision.  G.Z. and E.P. fabricated the tDBLG samples. R.K.K., L.A.P. and A.K.G. provided data on low-field magnetotransport in G/hBN superlattices. K.W. and T.T. provided the hBN crystals. S.S. and V.F. developed the kagom\'e network model, and S.S. performed the data analysis under V.F. supervision. A. G.-R. and V.F. modelled miniband spectra in tDBLG and G/hBN heterostructure. V.F., F.K.deV. and S.S. wrote the paper. All authors discussed the paper and commented on the manuscript.
All authors declare no competing interests. All data is available from the authors upon reasonable request.

\section{Supplementary Materials include:}
Materials and Methods:
\begin{itemize}
    \item tDBLG device characterization and experimental data analysis
    \item G/hBN device characterization and data analysis
    \item Maslov and Berry phases
    \item Shortest interfering paths
    \item Details of saddle-point calculation, leading to Eq. (4)
    \item Calculation of dispersion for tDBLG and G/hBN
    \item Temperature dependence  of kagom\'e oscillation amplitude and estimate of coherence length
\end{itemize}
Figs. S1 –-- S8

\bibliography{references}

\clearpage
\section*{Supplemental Material for "Kagom\'e quantum  oscillations in graphene superlattices"}
\setcounter{figure}{0}
\setcounter{page}{0}
\setcounter{equation}{0}
\renewcommand{\thefigure}{S\arabic{figure}}
\renewcommand{\theequation}{S\arabic{equation}}
\renewcommand\thesection{S\arabic{section}}

\section{tDBLG device characterization and experimental data analysis}
For this work we used the same device as reported on in Ref.~\cite{Tomic2021}. Details on the fabrication process, as well as optical images of the Van der Waals stack can be found there.
We measured $V_\mathrm{xx}$ and $V_\mathrm{xy}$, convert this to $\rho_\mathrm{xx}$ and $\rho_\mathrm{xy}$ using the applied current $I$ and the width of the mesa $W=\SI{2374}{nm}$ and the average distance between the contacts of $L=\SI{910}{nm}$, and  obtain $\sigma_\mathrm{xx}$ and $\sigma_\mathrm{xy}$ through tensor inversion.
We control the temperature in a pumped Helium-4 cryostat by a heater and a feedback loop, allowing us to reach stable values from $\SI{1.2}{K}$ up to $\SI{60}{K}$.

\begin{widetext}
\begin{figure*}[h]
\centering
\includegraphics[width=\textwidth]{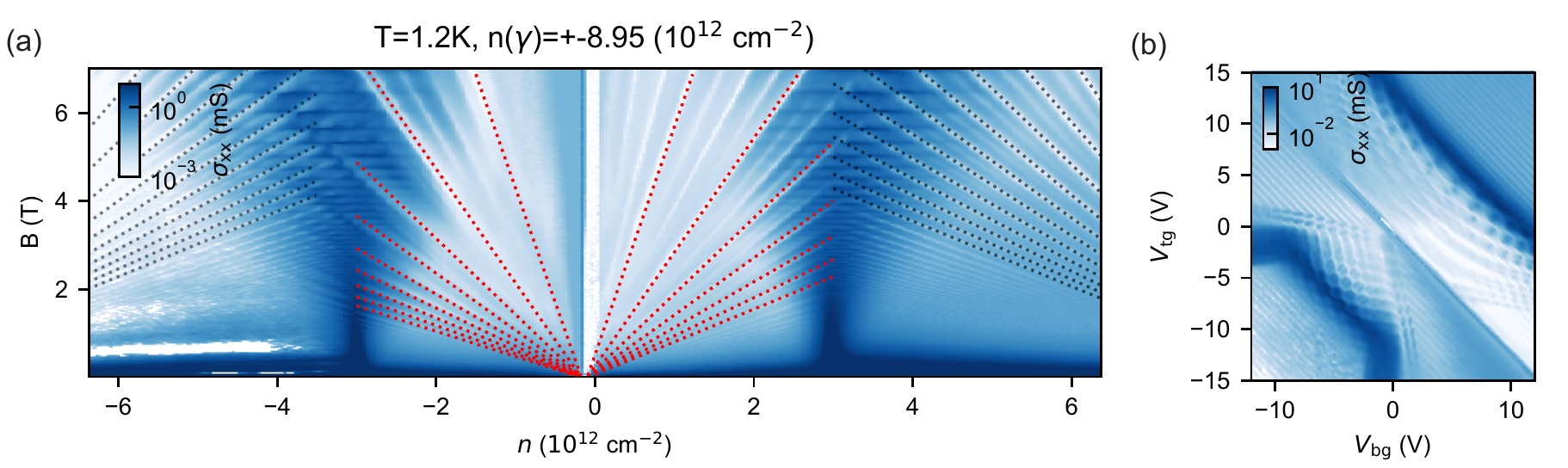}
\caption{(a) Shubnikov--de Haas map, $\sigmaxx(n,B)$ at zero displacement field, as plotted before in Fig.1(a). Here the SdH oscillations are fit (red and black dashed lines) to extract the degeneracies and full filling density $n(\gamma)$ as indicated.
(b) $\sigmaxx$ versus both bottom gate ($V_\mathrm{bg}$) and top gate ($V_\mathrm{tg}$) voltages at a constant magnetic field of $B=\SI{2}{T}$. Shubnikov--de Haas oscillations indicate constant density lines for the two bilayers.}
\label{suppfig:LLfits}
\end{figure*}
\end{widetext}

We calculate the density (in cm$^{-2}$) and displacement field (in V/nm) using the following equations:
\begin{equation}
    n = \frac{1}{e}(C_\mathrm{bg} V_\mathrm{bg} + C_\mathrm{tg} V_\mathrm{tg}) \cdot 10^{-4} - n_\mathrm{offset},
\end{equation}
\begin{equation}
    D = -\frac{ 0.5}{ \epsilon_0} (C_\mathrm{tg} V_\mathrm{tg} - C_\mathrm{bg} V_\mathrm{bg}) \cdot 10^{-9}
\end{equation}
where $C_\mathrm{xx}$ are the capacitance to the respective gates, $e$ is the elementary charge, $\epsilon_0$ the vacuum permittivity, and $n_\text{offset}$ an offset in the density.
The capacitances are obtained in two steps.
First we estimate the capacitance per area between the gate and the TDBG using the parallel plate capacitor model. Input parameters are the thickness of the hBN and AlO$_x$ layers as well as their dielectric constants.
Then we finetune the capacitance found by measuring the Landau levels as a function of the estimated density, and fitting the Landau levels using $n=h/q \cdot B/\nu$ where q is either +e for holes or -e for electrons (Fig.~\ref{suppfig:LLfits}(a)).
To crosscheck the ratio of the capacitances to top and bottom gate we then perform a measurements of the SdHO at constant magnetic field and as a function of both gates ( Fig.~\ref{suppfig:LLfits}(b)). The slope of the line where the total density equals zero gives us the ratio of the capacitances.

Furthermore, the fit of the Shubnikov--de Haas oscillations in Fig.~\ref{suppfig:LLfits} allows us to check the degeneracies.  The red dashed lines correspond to SdHO with a degeneracy of $8$. The degeneracy corresponds to spin, valley and minivalley (or layer) degree of freedom. The data reveals that this degeneracy is broken at higher magnetic fields where three additional lines appear between each pair of red dashed lines. This corresponds to the situation where the spin- and valley degeneracy is lifted, while the mini-valley degeneracy is maintained even at $\SI{7}{T}$. 

Finally, the oscillations emerging from large densities (i.e. $\SI{8.95e12}{cm^{-2}}$) are four-fold degenerate, as indicated by the black dashed lines.
The change from eight-fold degenerate electrons (at $n>0$) to four-fold degenerate holes occurs at the Lifshitz transition.
These fits allow us to estimate the density at full band filling and from that calculate a twist angle of $\SI{1.9}{\degree}$. We cross-checked this with the Brown--Zak oscillations and were able to refine it to $\SI{1.94}{\degree}$.%$\twist$.

\begin{widetext}
\begin{figure*}
\centering
\includegraphics[width=1\textwidth]{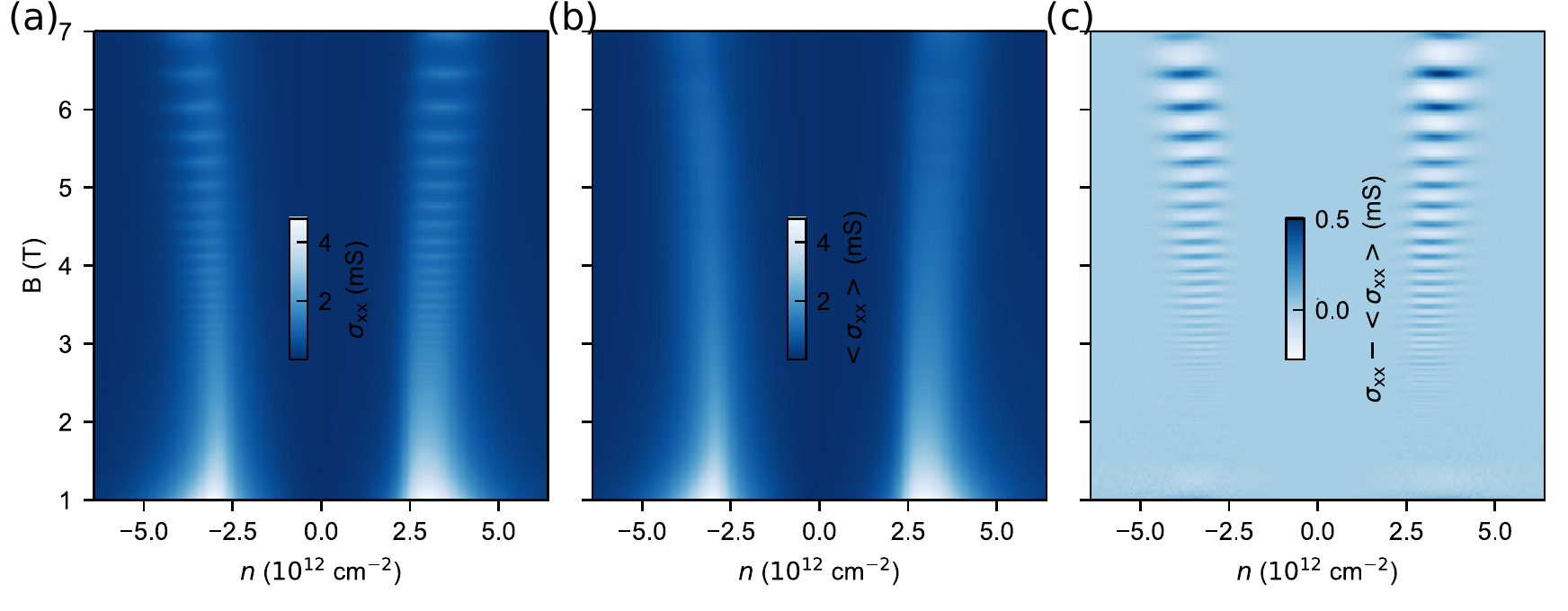}
\caption{Background subtraction used for Fig.~2(c). (a) obtained $\sigmaxx(n,B)$, (b) smoothed background $\langle \sigmaxx(n,B) \rangle$, (c) $\Delta\sigmaxx(n,B)$, the difference between (a) and (b).}
\label{suppfig:bg1}
\end{figure*}
\end{widetext}

%\subsection{Background subtraction and amplitude extraction}
The background subtraction procedure used to plot Fig.~3 %\ref{fig:fig3}
(a) in the main text is shown in Fig.~\ref{suppfig:bg1}. The kagom'e oscillations are visible in Fig.~\ref{suppfig:bg1}, where we show $\sigmaxx(n,q)$ at $\SI{20}{K}$. For better visibility, we apply a background subtraction procedure. We use a smoothing filter (Savitzky-Golay) along the $n$-axis (27 points, $1^{st}$ order) and obtain the background $\sigmaxxavg$ plotted in Fig.~\ref{suppfig:bg1}(b). By subtracting the two maps from each other, we obtain Fig.~\ref{suppfig:bg1}(c), i.e. $\Delta\sigmaxx=\sigmaxx-\sigmaxxavg$. The exact procedure is used for different temperatures for Fig.~3(e) and to obtain Fig.~3(d) smoothing is done over 19 points instead.

\begin{figure*}
\centering
\includegraphics[width=0.6667\textwidth]{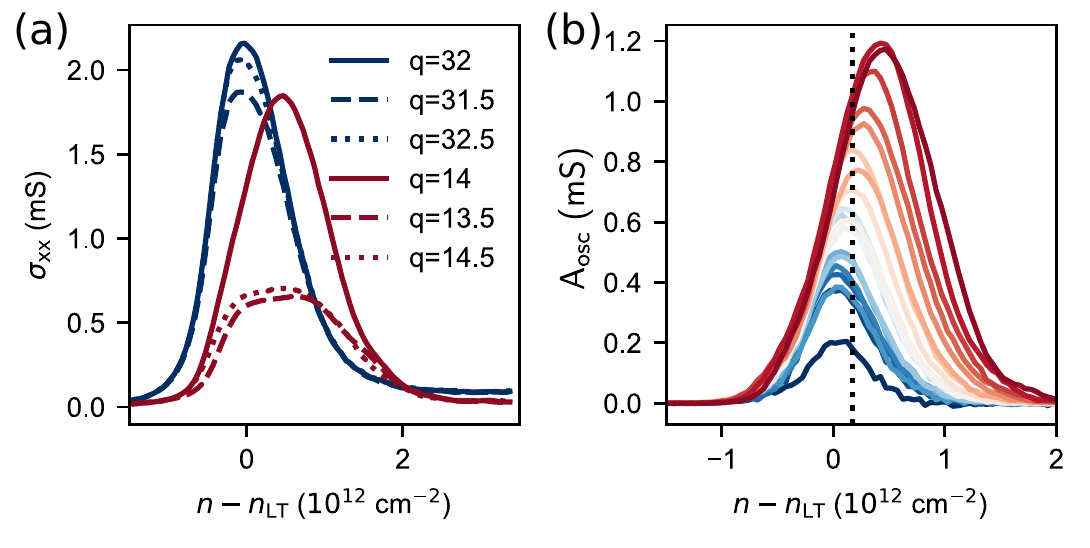}
\caption{(a) Examples of $\sigmaxx$ traces at half integer and integer $q$. (b) The extracted amplitude of the oscillations $\ABZ$ for $q$ ranging from 14 (red) to 32 (blue). The dashed line indicates the density at which the amplitude data for the temperature dependent measurements is taken. Both panels are for $T=\SI{15}{K}$.}
\label{suppfig:bg2}
\end{figure*}

Next, we obtain the kagom\'e oscillation amplitude $A_\text{osc}(n)$, used for Fig.~2(f). 
We first extract cuts $\sigmaxx(n)$ at integer $q$ and half-integer $q$, see Fig.~\ref{suppfig:bg2}(a). 
Then from these, we calculate $A_\text{osc}=\sigmaxx(q)-\nicefrac{1}{2}(\sigmaxx(q+\nicefrac{1}{2})+\sigmaxx(q-\nicefrac{1}{2}))$, i.e. $\ABZ$ is given by constructive minus destructive interference, see Fig.~\ref{suppfig:bg2}(b).
To obtain $\ABZ$ for many different temperatures for Fig.~2(f) we have taken linetraces at different temperatures at the density indicated in Fig.~\ref{suppfig:bg2}(b).

\section{G/hBN device characterization and data analysis}
The data for graphene aligned on hBN at $T=50$ K (Fig. 3d of main text) is a previously unpublished data, obtained in the experiment reported in Ref.\cite{Kumar2017}. The device details can be found in Supplementary Section 1 of Ref.\cite{Kumar2017}. The oscillatory part of $\sigma_{xx}$ has been extracted by the same method as for tDBLG sample discussed above. 

\section{Maslov and Berry phases} 
The phases of the amplitudes $\R$ and $\T$, presented in \eq{RT}, do not include the Maslov phase, which should be included in the amplitude as  $-i$ multiplier for each clock-wise $p_y$ turning point and as $i$ for each anti-clock-wise $p_y$ turning point \cite{MaslovTopological}.  Such separation of Maslov phases allows us to describe all the differently-oriented saddle points in a gauge-independent way.

A more precise version of semiclassical formalism  includes also the Berry curvature  and Berry magnetic moment corrections to the semiclassical phase, see~\cite{Wilkinson, Davis,Glazman18, GlazmanPRX}. Effectively, the Aharonov--Bohm interference correction that we describe corresponds to a phase gained on  a contour encircling the Brillouin zone.  The Berry magnetic moment can be accounted for by a magnetic field dependent shift of the dispersion, while the Berry curvature term leads to a Chern number of the band, $C_n$, which is integer and hence does not affect the phase of the kagom\'e oscillations.

\section{Shortest interfering paths}
\begin{figure*}[h]
\centering
\includegraphics[width=0.48\textwidth]{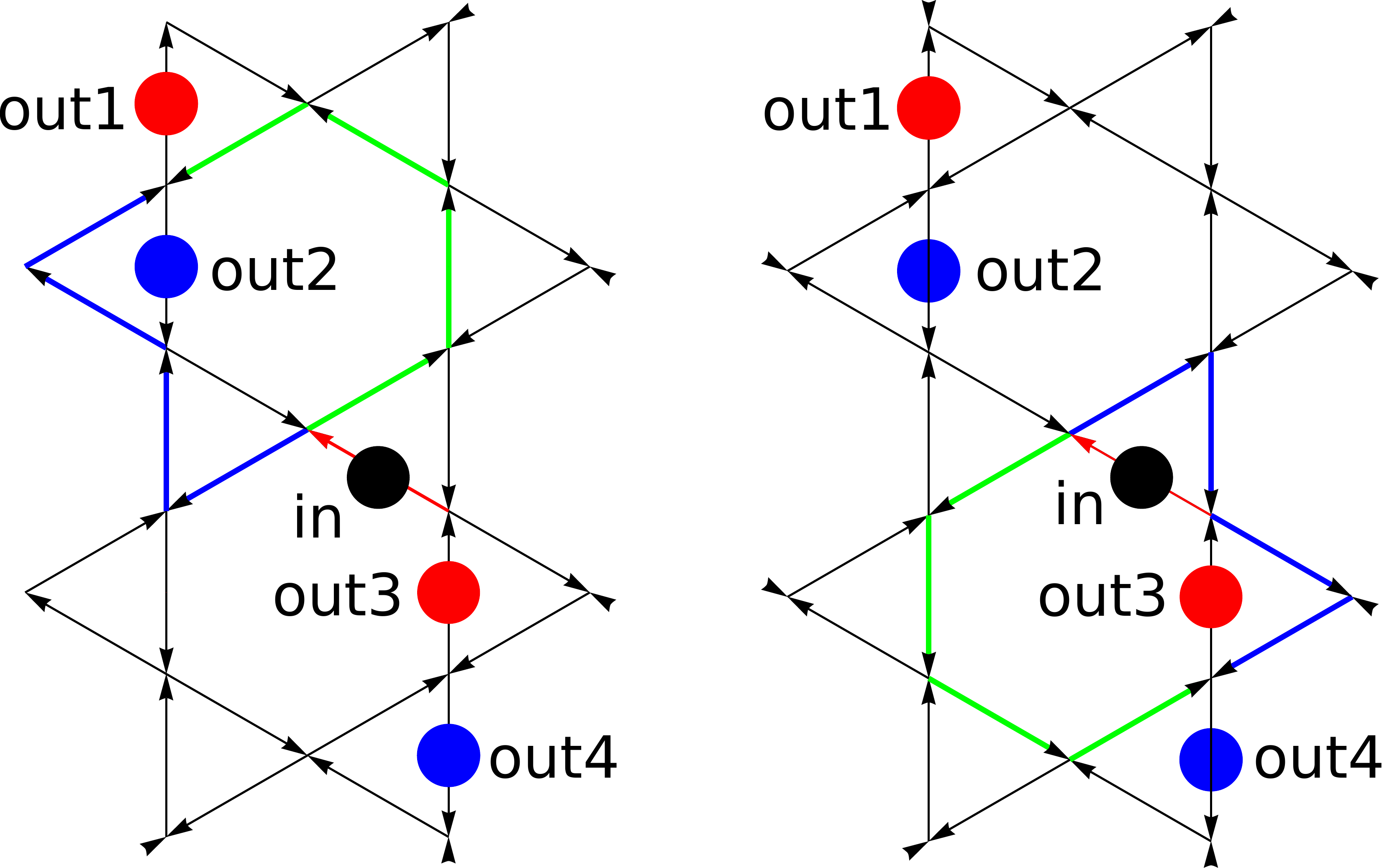}
\caption{Schematic illustration of 4 shortest-path amplitudes producing kagom\'e oscillations in conductance. Black dot is a starting point on a randomly selected segment, red dots indicate final segments for which probabilities get a  $+\cos(2 \pi \phi_0/\phi)$ interference correction (due to the real Maslov phase) and blue dots correspond to $-\cos(2 \pi \phi_0/\phi)$ terms (due to the imaginary Maslov phase). Interfering paths are shown in green and blue (for "out1" and "out2" on the left and for "out3" and "out4" on the right). 
Adding up the 4 leading contributions 
%as per \eq{Diffusion} S
shows that ``+'' terms dominate in the conductance, leading to $+\cos(2 \pi \phi_0/\phi)$ oscillations in the conductance.
}
\label{suppfig:Amplitudes}
\end{figure*}

In this section, we describe all the shortest interference paths contributing to kagom\'e oscillations. 
We start with a random phase space point "in", as shown in Fig.\ref{suppfig:Amplitudes}, and show all 4 of the possible shortest interfering contributions. For "out" points chosen on any other segments (not shown in Fig.\ref{suppfig:Amplitudes}), the Aharonov--Bohm interference contribution will contain longer path and will not be dominant in the limit of short phase coherence.   

Explicitly, the amplitudes are 
\begin{widetext}
\beqa
 \langle {\rm out1 |in }\rangle &=& \alpha_\text{diff} + e^{i \phi} \left[ \T^3 \R^2 + \T^3 \R^2 e^{i \frac{ e B (\cA_1+ \cA_2 + \cA_3)}{\hbar}} \right] e^{-\frac{\L_\text{in-out1}}{2 \l}} 
\\
\langle {\rm out2 |in }\rangle &=& \alpha_\text{diff} + e^{i \phi} \left[ -i \T^2 \R^3 + i \T^4 \R e^{i \frac{ e B (\cA_1+ \cA_2 + \cA_3)}{\hbar}} \right] e^{-\frac{\L_\text{in-out2}}{2 \l}} 
\\
\langle {\rm out3 |in }\rangle &=& \alpha_\text{diff} - e^{i \phi} \left[  \T \R^4 + \T^5  e^{i \frac{ e B (\cA_1+ \cA_2 + \cA_3)}{\hbar}} \right] e^{-\frac{\L_\text{in-out3}}{2 \l}} \\
\langle {\rm out4 |in }\rangle &=& \alpha_\text{diff} + e^{i \phi} \left[ -i \T^2 \R^3 + i \T^4 \R  e^{i \frac{ e B (\cA_1+ \cA_2 + \cA_3)}{\hbar}} \right] e^{-\frac{\L_\text{in-out4}}{2 \l}}
\eeqa
\end{widetext}
where we have taken into account the Maslov phases.
When module-squared to get the probability, all the 4 amplitudes presented above 
produce an oscillatory term $\sim \pm \T^6 \R^4 \cos(2 \pi \phi/\phi_0)$  and the sign is $+$ for ``out1'' and ``out4'' and $-$ for ``out2'' and ``out3''. 

There are two possible approaches to extract conductivity: first is the Einstein relation between conductivity and diffusion, which  implies that a contribution to the conductivity is proportional to the squared distance, $\langle x^2 \rangle$,  covered between the two scatterings.
Since the contributions with $+$ sign are seen to have higher values of $\langle x^2 \rangle$,   the overall sign of oscillating term is $+ \cos(2 \pi \phi_0/\phi) $.
Another approach to conductivity is based on Kubo formula and involves the velocity-velocity correlator. Since the positive (negative) contributions have velocities with positive (negative) projection onto the initial velocity, the total contribution is, clearly, positive. 

\section{Details of saddle-point calculation, leading to \eq{oscillation_main} 
%Eq.(4)
}
The amplitude of kagom\'e oscillations is clearly peaked near the 
LT since it involves a product of both $\R$ and $\T$ terms, the typical energy width of oscillation region is $\sim \hbar e B \r$. 
A distinguishing feature of hexagonal network is that the  maximum amplitude of oscillations is  shifted from the Lifshitz transition point ($\mu=0$) in the direction of a higher doping  (as shown in Fig.~2c and Fig.~3 a,d of the Main Text). 
This occurs because  higher power of $\T$ compared to $\R$ is involved in the oscillation amplitude, \Eq{amplitude}, $\left |\T^6 \R^4\right | = \frac{e^{-\pi \mu}}{32 (\cosh \pi \mu)^5}$. 
The maximum of this expression is shifted by  
\beq \label{exact_max}
\eps_\text{max} = -\frac{e B \r \ln (3/2) }{2 \pi \hbar} = - \frac{\r B}{\phi_0} \ln\frac32 
\eeq
from the energy of Lifshitz transition. 
The peak in  $\left |\T^6 \R^4\right |$ is further broadened by accounting for the finite temperature. To evaluate the integral analytically, we can expand the exponents in Taylor series either around the LT ($\eps = 0$), or around the maximum ($\eps_\text{max}$), or, one can approximate the integrand with a Gaussian according to a mean value and dispersion.  All these approaches lead to very similar results that differ numerically by a few percent. 
Taylor-expanding the exponent around the LT point and also approximating the $n_F'$ with a Gaussian gives  
\beqa 
&\int d\eps \, n_F'(\eps) |\T^6 \R^4| \propto \nn \\  
&\int \frac{d \eps}{T} \,\exp\left[-\frac{5 \pi^2\hbar^2}{2  e^2 \r^2 B^2}\left(\eps + \frac{ 2 \r B}{5 \phi_0} \right)^2 \right] \exp\left[\frac{-(\eps-\eps_F)^2}{4 T^2}\right],
\nn \eeqa
leading to \Eq{oscillation_main} of the Main Text. Note that the resulting position of the maximum in \Eq{oscillation_main}, $\eps_\text{max (saddle)} = -\frac{2}{5} \frac{\r B}{\phi_0}$ is numerically very close to the value in \eq{exact_max}, because $\ln(3/2) = 0.4055 \approx 2/5 $.  

\section{Calculation of dispersion for tDBLG and G/hBN}
Although our results do not depend on the details of dispersion,  we used the dispersion relations to plot the network and estimate the Gaussian curvature near the saddle-points.
The dispersion of tDBLG was calculated according to Refs.\cite{AitorRaman,AitorTrigonal}, and  the details can be found in Supplementary to Ref.~\cite{Tomic2021} (parameters of Slonczewski-Weiss-McClure model used are   $\gamma_0 = 3.16 \eV$, $\gamma_1 = 0.381 \eV$, $\gamma_3=-0.38 \eV$, $\gamma_4 = 0.14 \eV$).  For aligned G/hBN system, we used
Refs.~\cite{WallbankUmklapp,LeeWallbank16} (parameter values $U_0 = 8.5 \meV,\ U_1 = -17 \meV , \  U_3= -14.7 \meV$). 

\begin{figure*}
\centering
\includegraphics[width=0.6667\textwidth]{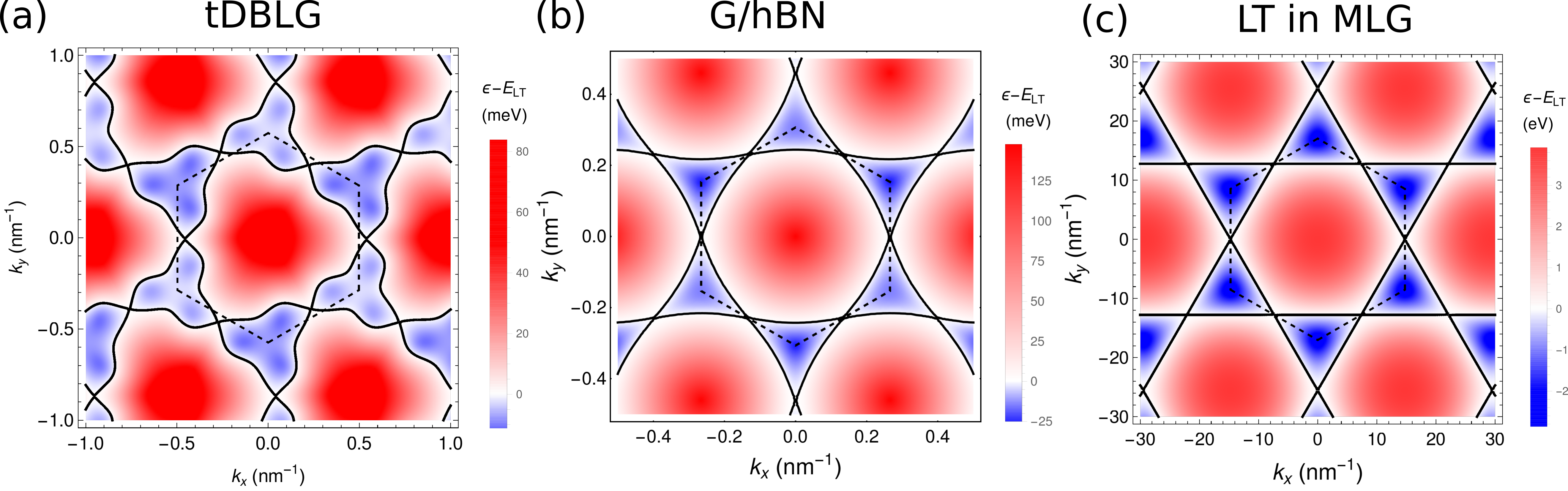}
\caption{Comparison of Fermi contours at LT  between (a) $1.9^\circ$ tDBLG (conductance band), (b) valence band of G/hBN, (c) graphene monolayer doped up to LT  In all the cases we see the same LT topology where one FS turns into two complimentary ones.}
\label{suppfig:kagome}
\end{figure*}

The kagom\'e networks of saddle-point trajectories in magnetic field look very similar for tGBLG and G/hBN and shown in Fig.\ref{suppfig:kagome}.

\section{Temperature dependence  of kagom\'e oscillation amplitude and estimate of coherence length} 
\begin{figure*}[h]
\centering
\includegraphics[width=1\textwidth]{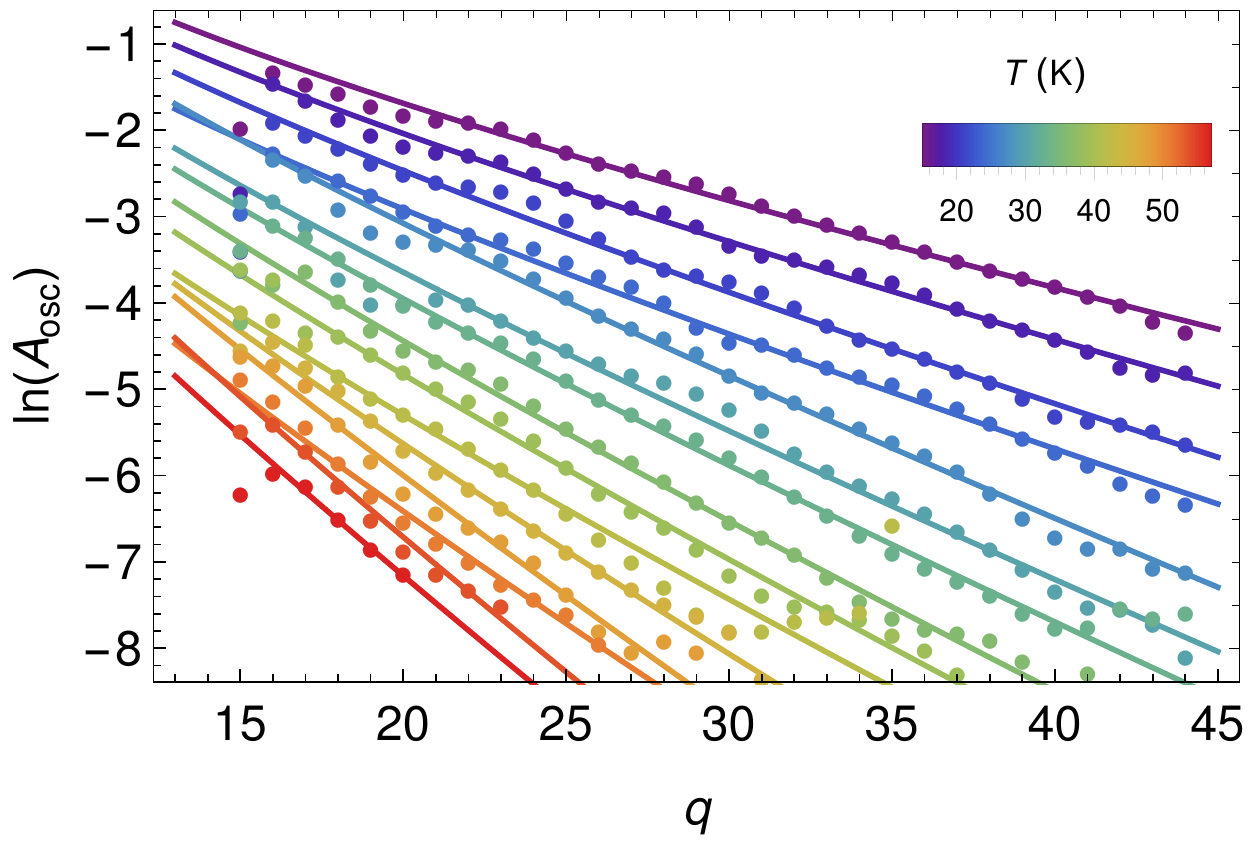}
\caption{Full data for $\ln \ABZ$ as a function of $q$ for different temperatures, the lines correspond to $\ln \ABZ =a_0 - \frac{2 \L}{\ell} - \ln(q)$ fit according to Eq.(4) of the main text (note that $\L(q) \approx \frac{\hbar}{e B}\, 2\nm^{-1} \propto q $, producing the main contribution to slope of the lines).
\label{suppfig:dephasing}}
\end{figure*}

%\eq{\label{oscillation_main}}}
In Fig.~\ref{suppfig:dephasing} we show the full dataset for oscillation amplitude at doping where it is maximal, $n\approx n_{max}$. 
The minimal real-space path required to see the oscillations can be estimated by taking the length of a green path in Fig.1 (same as the length of a blue path) in momentum space, and rescaling that into the real space length using a factor $1/(e B)$, resulting in $\L \approx \frac{\hbar}{e B}\, 2\nm^{-1}$. Note that $2 \L$ is comparable to the perimeter of an extended magnetic supercell, $\L_q = 4\sqrt{\frac{2}{\sqrt{3}}} q \sqrt{\cA_{\varhexagon}}= \frac{\hbar}{e B}\, 3.98 \nm^{-1}$, where $\cA_{\varhexagon}  \approx 46 \nm^2$ for the studied tDBLG.   

\begin{figure*}[h]
\centering
\includegraphics[width=1\textwidth]{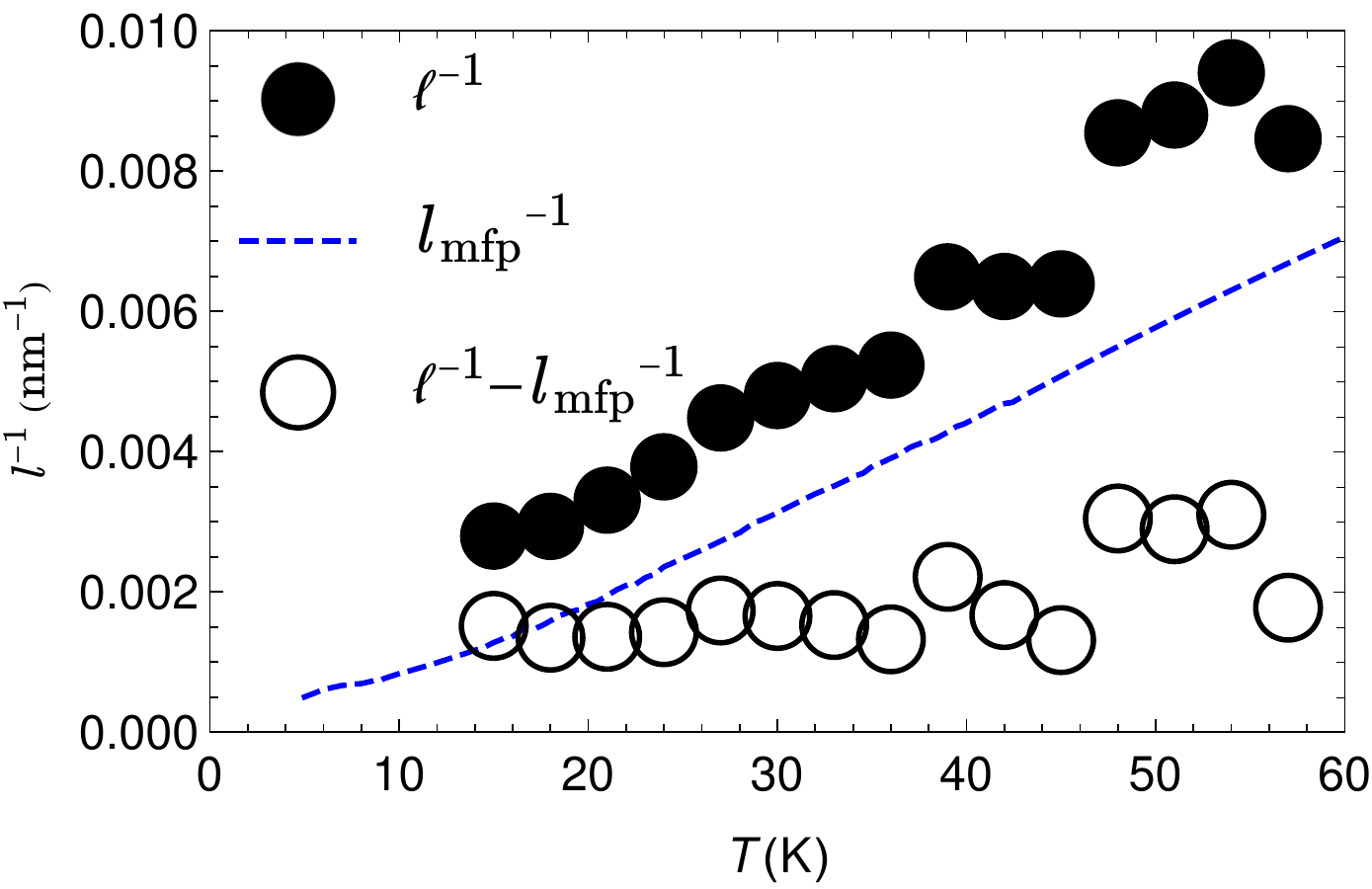}
\caption{Length $\ell$ vs the mean free path, $l_\text{mfp}$ extracted from conductivity. We plot $\ell^{-1}$,  $l_\text{mfp}^{-1}$ and the difference $\ell^{-1} -   l_\text{mfp}^{-1}$, showing weak temperature dependence of dephasing contributions.}
\label{suppfig:mfpPlot}
\end{figure*}

Fitting the measured magnetic field dependence of oscillations amplitudes with \eq{oscillation_main} and length $\L$, we extract the value of the scattering parameter, $\ell^{-1}(T)$, which determines the loss of  electrons from ballistic propagation. The result of such fitting are displayed as black circles in Fig.\ref{suppfig:mfpPlot}, where we also compare it with the inverse of the mean free path, $\ell^{-1}_\text{mfp}$ (blue curve), determined from the conductivity measured at the Lifshitz transition density and $B=0$ (here, we use average velocity estimated from the computed dispersion, shown in Fig. 3 of the main text). This comparison shows that momentum relaxation is slower than decay of ballistic beam, which could be expected, based on that $\ell^{-1}_\text{mfp}$ is more sensitive to the large momentum transfer upon scattering, whereas $\ell^{-1}$ is caused by both large and small momentum transfers. Both of these two quantities are temperature-dependent, indicating the contribution of inelastic scattering processes, most likely, generated by phonons. In contrast to that, their difference, shown in  Fig.\ref{suppfig:mfpPlot} for the overlapping temperature interval for the available data using empty circles, is almost temperature-independent, suggesting that the low-angle scattering in the system is mostly elastic, rather than inelastic.

\begin{figure*}[h]
\centering
\includegraphics[width=1\textwidth]{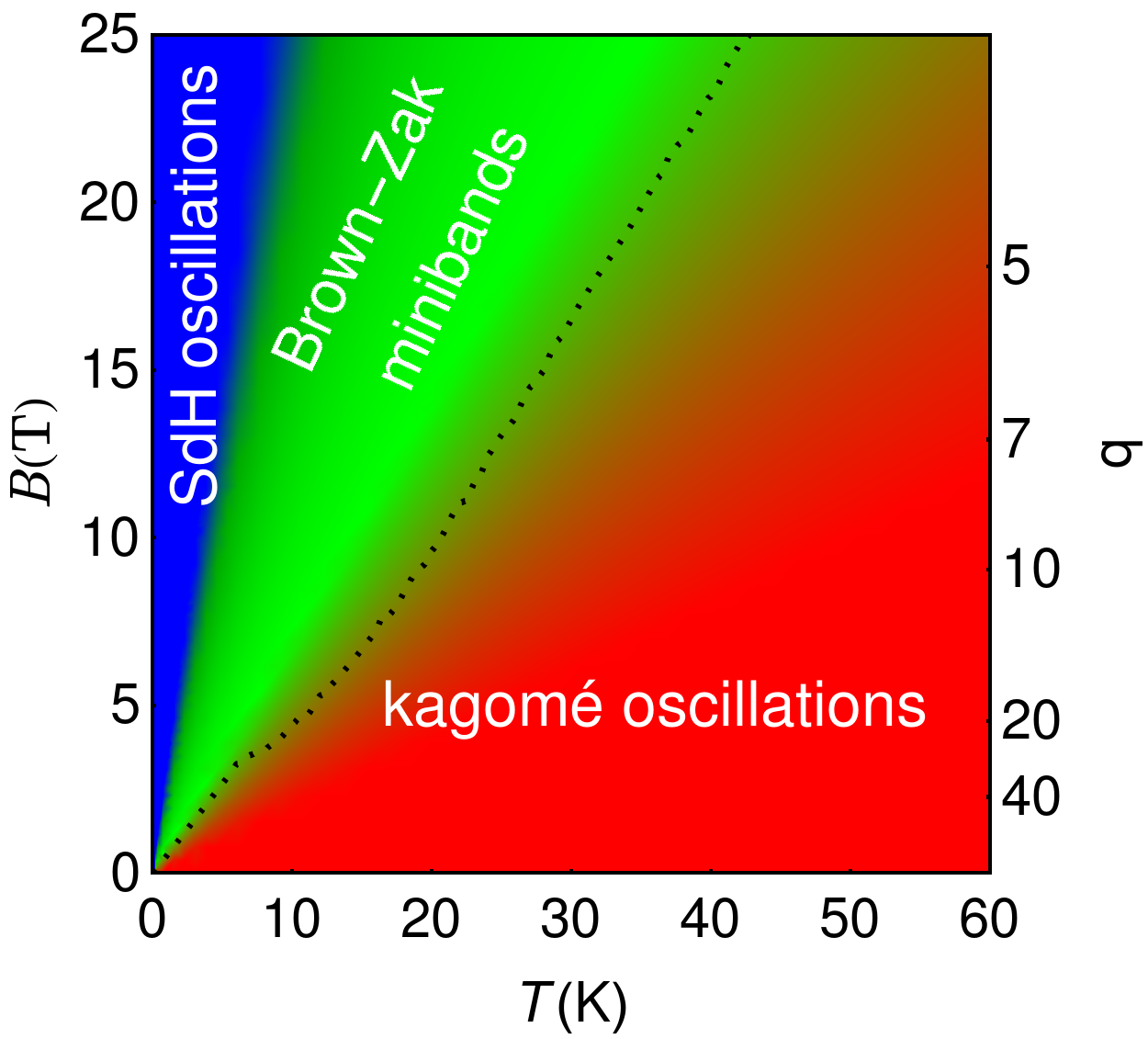}
\caption{Regions where different type of phenomena appear.  For $B<B^*$ the Brown--Zak minibands are not yet formed, still the kagom\'e oscillations are visible. For $B>B^*$, there are sizeable contributions from path encircling multiple magnetic unit cells, leading to formation of Brown--Zak minibands. }
\label{suppfig:Bstar}
\end{figure*}

Having compared the lengths $\ell_\text{mfp}$ and $\ell$, we come back to the discussion of their influence on   conditions for the formation of Brown--Zak magnetic minibands. We note that the latter are determined by Bragg scattering from the superlattice with a period inflated  $q$ times as compared to the underlying moir\'e pattern \cite{Brown1964,Zak1964,Chen2014}. Note that such magnetic supercell has a perimeter $\L_q \propto B^{-1}$. For establishing periodicity, required for Bragg scattering, the interfering electron waves should propagate ballistically over at least two unit cells, exploring their areas for establishing the magnetic flux commensurability with the flux quantum, so that the minimal condition for the formation of magnetic miniband spectrum would require that 
\begin{equation}
\ell_\text{mfp}(T)>2\L_q(B).
\label{boundary}
\end{equation}
This requirement is reflected on the $B-T$ parametric diagram in Fig. \ref{suppfig:Bstar}, as a borderline between the the 'higher' magnetic field regime, where magnetic minibands would have a chance to form (green area), and a low-field region (red area), where only kagom\'e oscillations appear, as a precursor of Brown--Zak minibands formation. By inspection, we notice that  the data displayed in Fig. 2 and discussed in Fig. 3a  belong to the latter parametric interval.

%\bibliography{references}

\end{document}